\documentstyle[11pt,aaspp4]{article}

\lefthead{Comastri A., et al.}
\righthead{X--ray spectra of Blazars}

\begin{document}

\title{On the soft X--ray spectra of $\gamma$--loud blazars}
\author{Andrea Comastri} 
\affil{Osservatorio Astronomico di Bologna}
\authoraddr{via Zamboni 33, I--40126, Bologna, Italy}
\authoremail{comastri@astbo3.bo.astro.it}
\author{Giovanni Fossati}
\affil{Scuola Internazionale Superiore di Studi Avanzati, Trieste}
\authoraddr{via Beirut 2--4, I--34014, Trieste, Italy}
\authoremail{fossati@sissa.it}
\author{Gabriele Ghisellini}
\affil{Osservatorio Astronomico di Merate}
\authoraddr{via Bianchi 46, I--22055, Merate, Italy}
\authoremail{gabriele@merate.mi.astro.it}
\and
\author{Silvano Molendi} 
\affil{Istituto di Fisica Cosmica e Tecnologie Relative del CNR, Milano}
\authoraddr{via Bassini 15, I--20133, Milano, Italy}
\authoremail{silvano@ifctr.mi.cnr.it}

\begin{abstract}

ROSAT observations of a large sample of bright $\gamma$--ray (E $>$ 100 MeV) 
blazars are presented.
Results of a detailed spectral analysis in the soft $\sim$ 0.1--2.0 keV
energy range are discussed in relation to the overall energy distribution
with particular emphasis on the relation between X--ray and $\gamma$--ray
properties.
A significant anti--correlation between X--ray and $\gamma$--ray spectral
shapes of flat radio spectrum quasars (FSRQ) and BL Lacs has been discovered. 
A different shape in the overall energy distributions from radio to 
$\gamma$--ray energies between FSRQ and BL Lacs is also implied by the 
correlation of their broad--band spectral indices $\alpha_{ro}$ and 
$\alpha_{x \gamma}$.
Both the above correlations can be explained if both the IR to UV emission 
and the hard X--ray to $\gamma$--ray emission originate from the same 
electron population, via, respectively, the synchrotron process and the 
inverse Compton mechanism.
We suggest that a key parameter for understanding the overall energy 
distributions of both classes of objects is the energy at which the 
synchrotron emission peaks in a $\nu-\nu F(\nu)$ representation.

\end{abstract}


\newpage   
\singlespace 
\section{Introduction}

The most important result of the CGRO--EGRET instrument in the field 
of extragalactic astronomy is the discovery that blazars
(i.e. Flat Spectrum Radio Quasars (FSRQ) and BL Lac objects)
emit most of their bolometric luminosity in the high $\gamma$--ray
(E $>$ 100 MeV) energy range.

At present more than 60 blazars detected by EGRET 
have been identified (Fichtel et al. 1994; von Montigny et al. 1995; 
Thompson et al. 1995; Nolan et al. 1996; Sreekumar et al. 1996; 
Lin et al. 1996; Dingus et al. 1996).
The majority are FSRQ while about a dozen are classified as BL Lac objects.
Furthermore, the Whipple observatory detected 3 BL Lac objects
above 300 GeV: Mkn 421 (Punch et al. 1992); Mkn 501 (Quinn et al. 1996) 
and PKS 2344+514 (Fegan 1996), the latter two not yet detected by EGRET.
\par
The origin of this high energy, strong and variable emission is still 
not clear. The very nature of the detected sources, all of them being
associated with blazars, strongly suggest that relativistic motion and 
beaming of the emitted radiation are required.

The blazars broad--band energy distribution (BBED) from radio to $\gamma$--ray 
energies, in a $\nu$--$\nu F_{\nu}$ representation can be characterized
by two peaks: one in the IR to UV band and the second in the MeV--GeV band.
The smooth rise from radio to infrared frequencies is followed 
by a cut-off in the optical-UV range with a minimum in the 
X--ray range. 
In some objects, in the X--ray band there is an upturn towards higher energies,
suggestive of a connection with the peak of the 
energy distribution in the $\gamma$--ray range.
\par
The observed BBED, previously modeled by Comastri, Molendi \& Ghisellini
(1995, hereinafter Paper I) for a small sample of BL Lac objects and
by Sambruna, Maraschi \& Urry (1996) for a larger sample of blazars,
suggests that the radio to the UV emission is produced 
by the synchrotron process, while the inverse Compton mechanism  
is responsible for the high energy $\gamma$--ray emission.
However, the origin of the seed photons and the location
and size of the emitting region(s) are unknown.
Seed photons can be produced internally to the emitting blob or jet
(by synchrotron emission, as in the jet model of Maraschi,
Ghisellini \& Celotti, 1992), or be produced externally, by
the accretion disk (as in the model of Dermer \& Schlickeiser, 1992),
by the broad line region illuminated by the disk (as in the model of
Sikora, Begelman \& Rees, 1994), and/or by some scattering material
surrounding the jet (Blandford \& Levinson, 1995).
Finally, the blob itself could illuminate a portion of the broad line
clouds, whose reprocessed line radiation can dominate the radiation
energy density in the blob (as in the model of Ghisellini \& Madau, 1996).

The main purpose of the present work is to study, for a sizeable sample 
of bright $\gamma$--ray sources, the role of the X--ray emission with 
respect to the BBED, with particular emphasis on the relation between 
the X and $\gamma$--rays.
\par
The soft X--ray band lies where both synchrotron and self--Compton 
components are expected to provide a significant contribution
to the total emission. 
For most of the sources, the X--ray flux density allows 
to estimate the minimum of the BBED, and hence to derive 
physical parameters to constrain the theoretical models 
(synchrotron--self--Compton, hereinafter SSC, 
Comptonization of ambient radiation field).
Further constraints on the emission mechanisms can be obtained 
from the variability pattern and the slope of the X--ray spectrum.
In fact a steep X--ray spectrum is expected if the emission is mainly
due to the synchrotron process; in this case the X--ray data coupled with 
the radio to UV spectrum allow the measurements of the peak of the BBED and 
the determination of the physical parameters of the synchrotron emission. 
Rapid flux and spectral variability is expected in this case.
Viceversa, a flat X--ray spectrum, expected if the emission
is dominated by the Compton flux, suggests that the same process
is producing both the X-- and $\gamma$--ray components, which therefore
should show correlated variability patterns.

The outline of the paper is as follows: 
in Section 2 the sample selection is briefly presented; in Section 3 the 
analysis of all the available PSPC observations of $\gamma$--ray blazars 
is described in some detail; in Section 4 the correlations among the derived 
X--ray spectral properties and other bands are illustrated; in Section 5 
the origin of the $\gamma$--ray emission and its connection with the X--rays 
is discussed; a summary of the main conclusions is presented in Section 6.

\section{The sample}

A list of $\gamma$--ray blazars has been compiled collecting all the data 
available in the literature 
up to September 1996 and contains about 60 sources.
The $\gamma$--ray bright source list has been cross--correlated with the 
ROSAT PSPC pointed observations public archive.   
About half (27) of the sources were either the target of pointed 
observations or in the field of view (within about 40 arcmin) 
of a different target. 
For 19 of the sources, which were not observed in pointing mode, 
ROSAT All Sky Survey (RASS) data were available in the literature 
(Brinkmann, Siebert \& Boller 1994; Brinkmann et al. 1995). 
Finally we have included 7 more sources with X--ray fluxes 
retrieved from missions other than ROSAT.

This selection yielded  53 bright $\gamma$--ray blazars which are listed
in Table 1. We give the IAU name (col. 1) and other names (col. 2) 
of the sources, their classification (col. 3), redshift (col. 4),
significance of the EGRET detection (col. 5) of the chosen viewing period
(col. 6), the reference for the $\gamma$--ray data (col. 7), the type of the
X--ray observation (col. 8) and its reference (col. 9).
A concern regards the selection of the $\gamma$--ray data, given the large 
amount of available observations and analyses of the same data 
by different authors. For most of the objects there exist several 
EGRET observations and/or for some of these more than one analysis is
available.
In these cases we have chosen the $\gamma$--ray flux and spectral slope 
according to the following criteria:
\par
i) spectral index and flux referring to the same observation, ii)
data corresponding to a single observing period, iii) when the same
data are analyzed by different authors, the results of the most 
recent analysis are preferred.

Even though not complete, this sample includes about 80 \% of all known
$\gamma$--ray blazars and therefore we feel justified in considering it
representative of the entire population.

\section{ROSAT Observations and data analysis}

\subsection{Data reduction}

The data reported here are from observations carried out
with the ROSAT X--ray telescope (Tr\"umper 1983)
with the PSPC in the focal plane (Pfeffermann et al. 1986).
The PSPC provides a bandpass in the range $\sim$ 0.1--2.4 keV
over a $\sim$ 2$^\circ$ diameter field of view, with a moderate
spectral resolution.
Most of the observations have been performed in the ``wobble" mode. 
In this mode the detector is moved from the pointing direction with 
a period of $\sim$ 400 s and an amplitude of $\sim$ 3 arcmin. 
In this way the effect of shadowing of the X--ray sources by the 
detector window structures like struts and wire mesh is minimized.
\par
The spatial profiles of the images for all the observations
in the $\simeq$ 0.1--2.4 keV range are consistent with those of point
sources convolved with the PSPC point spread function (PSF) according 
to the source spectral properties and keeping into account the background 
level. Source spectra were extracted from circular regions
with radii large enough (from 2 to 4 arcmin) to ensure
that all the soft counts be included, given the electronic 
ghost imaging which widens the PSF  below $\sim$ 0.2--0.3 keV (Hasinger 
et al. 1992).
Background spectra were taken either from annulii centered on the sources
or from circular regions uncontaminated by nearby sources with extraction
radii as large as 10 arcmin. 
The large extraction cells ensure a good statistics for the
modelling of the background spectrum and allow to average
the background small scale spatial fluctuations. 
Different background regions for each source have been extracted 
and compared. In all the cases the background was stable without 
any appreciable variability within the statistical errors.
Corrections were included for vignetting, especially 
important for the off--axis sources, and PSPC dead time.
The ROSAT observation log is shown in Table 2.
The reported count rates are in pulse invariant (PI) channels 
and are background subtracted. The lowest channel used was 11, 
as the detector response matrix is not calibrated below this channel,
the highest channel depends on the source spectrum and the signal--to--noise 
ratio of the observation, ranging from channel 200 to 240.
A major concern in the analysis of X--ray spectra 
is the uncertainty in the PSPC response matrix calibration.
Actually two matrices have been officially released 
in order to take into account the $\sim$ 60 Volts gain change 
applied on October 14, 1991.
The first matrix (MPE No. 06) has been adopted for the observations
carried out before this date (PV and AO1 targets), while the
latest release (MPE No. 36) has been used for all the
other observations.
The photon event files were analysed using the EXSAS/MIDAS 
software (version 94NOV, Zimmermann et al. 1993) and the extracted 
spectra were analysed using version 9.0 of XSPEC (Shafer et al. 1991).
\par
For some of the sources in our sample X--ray data have been 
previously published by different authors. In order to ensure
a uniform procedure for all the sample we have re--analysed 
all of them obtaining consistent results.
For two of the sources in our sample (1226+023$\equiv$3C 273 and 
1253--055$\equiv$3C 279)
several tens of pointed observations are available in the public archive 
as they were the subject of simultaneous multiwavelength campaigns.
In this case we have chosen only one (the longest) pointing in order 
to derive the spectral parameters.

\subsection{Spectral Analysis}

All the 27 sources have been clearly detected with
enough counts to allow spectral analysis.
The source spectra were rebinned in order to obtain a significant
S/N ($>5$) for each bin and fitted with a single power--law model 
($F_{\nu} \propto \nu^{-\alpha}$ where $\alpha$ is the energy spectral index)
plus absorption arising from cold material with solar
abundances (Morrison \& McCammon 1983).
The derived spectral parameters are given in Table 3, where the reported
errors are at 90 per cent confidence level
(Lampton, Margon \& Bowyer 1976). 
All the spectra were fitted with i) the column density 
fixed at the Galactic value, ii) and free to vary. 
Accurate values for the Galactic column
densities towards most of the objects in our sample 
have been retrieved from 21 cm radio surveys 
(Elvis, Lockman \& Wilkes 1989; Lockman \& Savage 1995; 
Murphy et al. 1996).
The typical error in such measurements is estimated to be of the
order of $\simeq$ $1-3 \times 10^{19}$ cm$^{-2}$. 
For the remaining objects the Galactic column density values are
from Dickey \& Lockman (1990). The errors in this case are of the 
order of $\simeq$ $1 \times 10^{20}$ cm$^{-2}$.
\par
It should be noted that radio observations only detect interstellar
atomic hydrogen and not molecular gas (i.e. H$_2$, CO, etc.).
Therefore the total effective absorbing column density could be 
underestimated by 21 cm measurements.
Given this possibility we checked for Galactic CO emission in the
direction of all the blazars in our sample.
Carbon monoxide emission has been detected in the direction
of two of the objects in our sample namely:
0528+134 and 2251+158 (Liszt \& Wilson 1993; Liszt 1994).
Accurate measurements of the CO profile intensity have been recently 
obtained by Liszt (1996; private communication) as 2.15 K km/s for 0528+134
and 0.77 K km/s for 2251+158.

The molecular hydrogen column density depends on the assumed 
CO--to--H$_2$ conversion factor. 
For 0528+134 a rather conservative assumption of $3 \times 10^{20}$ 
molecules cm$^{-2}$ K$^{-1}$ km$^{-1}$ s (Strong et al. 1988) yields
an equivalent molecular hydrogen column density N(H$_2$) $\simeq 1.3 \times
10^{21}$ cm$^{-2}$, which combined with the atomic value of
$2.8 \times 10^{21}$ cm$^{-2}$ (Dickey \& Lockman 1990) 
gives a total absorbing column of $\sim 4.1 \times 10^{21}$ cm$^{-2}$.

Given the relatively high Galactic latitude of 2251+158 
($b II \simeq -40\arcdeg$) a much lower
conversion factor of $0.5 \times 10^{20}$ molecules 
cm$^{-2}$ K$^{-1}$ km$^{-1}$ s has been assumed 
(see De Vries, Heithausen \& Thaddeus 1987).
The molecular hydrogen column density of $\simeq 0.77 \times 
10^{20}$ cm$^{-2}$, combined with the atomic value of 
$7.06 \times 10^{20}$ cm$^{-2}$ (Murphy et al. 1996) 
gives a total absorbing column of $7.83 \times 10^{20}$ cm$^{-2}$.

In most cases a single power law spectrum with the absorption fixed at 
the Galactic value provides an excellent description of the data, 
while for some of the objects either the fits are statistically
unacceptable and/or the column density inferred from the fit
is not consistent with 
the Galactic value, even considering the errors on the Galactic $N_H$ 
(see column 6 in Table 3), 
thus requiring a more complex description of the spectral shape.

A double--power--law model has been fitted to all these objects 
and the results are reported in Table 4.
We point out that for all of the objects the improvement with respect 
to the single power law fit with $N_H \equiv N^{Gal}_H$ is statistically
significant, as indicated by the F--test, at $>$ 99.9\% level.

On the basis of the analysis presented in this subsection it appears 
that a two component (i.e. a double--power--law) model is required to 
describe the soft X--ray
spectrum for some of the objects, usually the ones with 
the highest counting statistics in the sample.
It is suggested that deeper and/or broader band ($\sim$ 0.1--10 keV)
observations will reveal spectral complexity for other objects.

\subsection {Statistical Analysis}

The range of measured values of the energy spectral indices is large,
going from $\simeq$--0.3 to 2.3. 
We find a significantly different distribution of spectral indices between 
BL Lacs and FSRQ, as can be clearly seen from the histogram shown 
in Figure 1.
The mean spectral properties have been computed assuming the slopes
derived with $N_H \equiv N^{Gal}_H$ if the X--ray absorption is 
consistent with the Galactic value (col. 3 in Table 3), 
while the slopes derived from
the fit with $N_H$ free have been adopted in the other cases (col. 7 in 
Table 3). This choice allows a better estimate of the continuum shape
for the objects showing either intrinsic absorption or soft excess emission
(cfr. Table 4). 
We have also included the FSRQ 1156+295 using the 2--10 keV {\it Ginga}
slope (Lawson \& Turner 1996).  
When more than one observation was available (see Table 2) a weighted mean 
spectral slope for each object has been adopted. 

A method for estimating the mean and the intrinsic dispersion 
of the parent population is the maximum--likelihood algorithm 
(see Maccacaro et al. 1988).
The algorithm assumes that the intrinsic distribution of energy indices
is Gaussian, and allows the determination  of the mean 
$\alpha$, the intrinsic dispersion $\sigma$, and the respective errors,
weighting the individual energy indices according to their measured errors.
The results are as follows:

$$ 
{\rm BL~Lacs:}\qquad
\langle \alpha_X \rangle  = 1.43 \pm 0.21 ;\quad \langle \sigma_{\alpha_X}
\rangle = 0.43^{+0.22}_{-0.12} 
$$

$${\rm FSRQ:~~}\qquad \langle \alpha_X \rangle = 0.67 \pm 0.13 ; \quad   
 \langle \sigma_{\alpha_X} \rangle
= 0.29^{+0.14}_{-0.07} 
$$

where the confidence intervals are at the 90 per cent confidence level
for one interesting parameter.

In order to test if the obtained values are dominated by the brightest
sources a simple unweighted mean has been computed.
The obtained values: $\langle \alpha_x 
\rangle = 0.63$, $\sigma$ = 0.23 for the 16 FSRQ, and 
$\langle \alpha_x \rangle = 1.43$, $\sigma$ = 0.50 for the 12 
BL Lacs objects, where $\sigma$ is the dispersion on the mean index, 
are consistent with the maximum likelihood analysis.

The two means are significantly different at $>$ 99.99\% level 
according to a Student t--test and taking into account the unequal variances
of the two populations.  
A Kolmogorov--Sminorv (K--S) test for the distribution of the observed values
gives almost identical results for the significance of the difference 
between the two populations ($>$ 99.99 \%).

The obtained results indicate that the X--ray spectra of BL Lacs objects
are steeper than FSRQ as first suggested by Worral \& Wilkes (1990),
and that the intrinsic distribution of spectral indices 
is not consistent with a single value.

Among the BL Lac objects present in our list, 4 are classified as
X--ray type BL Lacs (1101+384, 1652+398, 2005--489 and 2155-304),
according to their radio to X--ray flux ratio 
(Padovani \& Giommi 1995). 
The mean X--ray slope for these 4 objects is 1.75$\pm$0.33, while
for the remaining 8, radio--selected type BL Lacs, we find 
$\langle \alpha_X \rangle =1.26\pm 0.46$. Despite the low number of X--ray
type BL Lacs, the two slopes are different at $>$ 96.6 \% 
as indicated by the Student t--test, in 
agreement with our previous results (Paper I).

\subsection {Temporal analysis} 

The background--subtracted light curves have been computed 
for each source. As suggested by the wobble period of the telescope
the collected photons were binned in 400 or 800 s
time intervals depending on the source counting statistic.
The binned light curves for each source were fitted with a 
constant value. The probabilities associated to the $\chi^2$ 
statistic suggest that, in most of the cases, the light curves are not 
variable over the full observation with maximum deviations 
of the order of 20--30 \%.
Given that a detailed study of relatively small amplitude ($<$ 50 \%)  
variability is extremely difficult for sources observed on--axis 
due to the source obscuration from the fine wires mesh of the PSPC 
window (Hasinger, private communication), variations of this order
could be entirely due to instrumental effects.

Large flux variability on short timescales ($<$ 1 day)
has been detected only in S5 0716+714 which is discussed in a separate 
paper (Cappi et al. 1994).
Variations within the same PSPC observation are present 
in two more objects: 0219+428 $\equiv$ 3C 66 A and PKS 2155--304.

For 0219+428 the count rate decreased by about a factor 2 
on a timescale of $\sim$ 3.5 days without any significant spectral variability.
In the bright BL Lac object PKS 2155--304 an increase of 
a factor $\sim$ 1.7--1.8 on a timescale of $\sim$ 2.4 days has been 
detected. Given the high number of accumulated counts a time 
resolved spectral analysis was possible.
From an inspection of the light curve two states can be clearly identified: 
a low state at the beginning of the observation with a mean count rate of
$\sim$ 28 counts s$^{-1}$ and a high state ($\sim$ 48 counts s$^{-1}$)
at the end of the observation. In both cases a single power law with 
absorption fixed at the Galactic value
provide a good description of the observed spectrum, which flattens
from $\alpha = 1.42 \pm 0.05$ in the low state to $\alpha = 1.28 \pm 0.02$ 
in the high state, with a behaviour typical of BL Lac objects.

Finally, large amplitude (up to a factor 3) flux variability 
has been observed for some of the sources in the sample for which 
several observations were available in the public archive (Table 2).
The observed timescales range from a few days to several months.
For two objects (0851+202 $\equiv$ OJ 287 and  2005--489) 
spectral variability has been detected as well.
For a  detailed discussion of the spectral variability
in PKS 2005$-$489 we refer to Sambruna et al. (1995).
For OJ 287 a spectral steepening ($\Delta \alpha \simeq 0.4-0.5$) 
between the first two observations separated by about 7 months, 
is associated with a doubling count rate. 
A further observation, after two years, reveals a count rate 
comparable to the second observation, but a slope similar to the first.

\section{$\gamma$--ray and overall spectral properties}

\subsection{$\gamma$--ray energy distribution}

In Table 5 we list the $\gamma$--ray fluxes for all the sources in our sample,
together with spectral indices if available.
With the same maximum--likelihood algorithm applied in Section 3.3
we have computed 
the mean $\gamma$--ray slope for BL Lacs (10 out of 16 objects)
and FSRQ (30 out of 37 objects); the results are as follows: 

$$ 
{\rm BL~Lacs:}\qquad
\langle \alpha_\gamma \rangle = 0.87 \pm 0.13 ; \quad  \langle 
\sigma_{\alpha_\gamma} \rangle  < 0.26
$$ 

$$ 
{\rm FSRQ:~~~}
\langle \alpha_\gamma \rangle = 1.25 \pm 0.10 ; \quad \langle 
\sigma_{\alpha_\gamma} \rangle = 
0.25^{+0.10}_{-0.07} 
$$ 

where the confidence intervals are at 90 per cent level for one interesting 
parameter. 
The fact that the intrinsic dispersion of BL Lac spectral indices is consistent
with zero is probably due to the small number of objects.

A simple unweighted mean returns almost identical values:
$\langle \alpha_{\gamma} \rangle = 1.29$, $\sigma$ = 0.34 for FSRQ, and 
$\langle \alpha_{\gamma} \rangle = 0.87$, $\sigma$ = 0.32 for BL Lacs 
where $\sigma$ is the dispersion on the mean index.

The two means are significantly different at $>$ 99.7\% level 
according to a Student t--test. 
Similar results ($>$ 99.8 \%) for the significance of the difference 
between the two populations are returned from a K--S test on the 
distribution of the observed values.

The distributions of spectral indices indicate an opposite behaviour at 
$\gamma$--ray energies with respect to the X--ray band (Figure 1). 
For the subsample of 23 objects for which both X-- and $\gamma$--ray 
spectral slopes are available, 
a significant, at $>$ 98.6\% level using a non--parametric
Spearman rank test, anti--correlation has been found between 
the two spectral indices (Figure 2) when both FSRQ and BL Lacs are considered.
The correlation is however much weaker considering FSRQ only 
(82 \%) and is not present for BL Lacs only (15 \%).
FSRQ and BL Lacs occupy two different regions
in the $\alpha_x$--$\alpha_{\gamma}$ plane.
A flat X--ray spectrum is thus associated with a steep $\gamma$--ray 
slope for FSRQ while the reverse is true for BL Lacs.
A similar correlation was found by Wang, Luo \& Xie (1996) with a smaller 
sample (10 objects) and inhomogeneus X--ray data.

These findings suggest that either the $\alpha_x$--$\alpha_{\gamma}$ 
correlation is just the result of a fortituous proximity of the locii
occupied by BL Lacs and FSRQ in the $\alpha_x$--$\alpha_{\gamma}$ plane,
or the correlation points to an underlying physical mechanism
operating in both classes of objects.

\subsection{Broad--band energy distribution}

Radio and optical data for all the 53 objects in our sample have been 
collected from the literature in order to characterize the BBED
(Table 5). Since the $\gamma$--ray detection is likely 
to be associated with a high state of the source we have chosen
to list the brightest radio and optical fluxes, the latter corrected for 
reddening.

We have computed the broad band spectral indices $\alpha_{12}$ 
defined as $- log (L_2/L_1)/ log (\nu_2/\nu_1)$ where $L_1$ and
$L_2$ are the rest--frame luminosities observed at the frequencies 
$\nu_1$ and $\nu_2$, where $\nu_R$ = 5 GHz, $\nu_O$ = 5500 \AA, 
$\nu_x$ = 1 keV and $\nu_{\gamma}$ = 100 MeV.
The K--correction has been applied using the listed values of 
$\alpha_x$ and $\alpha_{\gamma}$ whenever available and the mean spectral 
indices for BL Lacs and FSRQ derived in the previous sections 
(cfr. $\S$ 3.2.1 and $\S$ 4.1) for sources with unknown spectra. 
Finally, $\alpha_R$ = 0 and $\alpha_O$ = 1 have been assumed 
at radio and optical frequencies. 

A highly significant anti--correlation, (at $>$ 99.99 \% using a
non--parametric Spearman rank test), 
has been found between $\alpha_{ro}$ and $\alpha_{x \gamma}$ (Figure 3),
when considering the entire class of blazars. 
A similar level of significativity is present applying a 
bi--dimensional K--S test (see Press et al. 1992). 
The correlation persists for each class of objects, although
with less significance (99.3\% for BL Lacs and 96.9\% for FSRQ).

Moreover, BL Lacs and FSRQs occupy different regions of the 
$\alpha_{ro}$--$\alpha_{x \gamma}$ plane.  
A relatively steep radio to optical spectral index implies a flat X to 
$\gamma$--ray flux ratio and is common among FSRQ, while the opposite is true 
for BL Lac objects. The average values are as follow:

$$ 
{\rm BL~Lacs:}\qquad
\langle \alpha_{ro} \rangle  = 0.47 \pm 0.12 ; \quad
\langle \alpha_{x \gamma} \rangle = 0.83 \pm 0.18 
$$ 

$$ {\rm FSRQ:~~~}
\langle \alpha_{ro} \rangle  = 0.69 \pm 0.10 ; \quad
\langle \alpha_{x \gamma} \rangle = 0.58 \pm 0.12 
$$

Even if the broad band indices $\alpha_{ro}$ and 
$\alpha_{x \gamma}$ are independent quantities, so that the effects
of spurious correlations are minimized, the observed correlation 
could be induced by other variables such as redshift and/or luminosity.

Indeed significant correlations (at $>$ 99.9 \%) have been found between 
both $\alpha_{x \gamma}$ and $\alpha_{ro}$ with redshift 
and $\gamma$--ray luminosity.
This is not surprising given that BL Lac objects have, on average,
lower luminosities and redshifts, compared to FSRQ.

The dependence on redshift and luminosity in the
$\alpha_{ro}$--$\alpha_{x \gamma}$ correlation can be removed 
applying a partial correlation analysis (Kendall \& Stuart 1979).
The results indicate that excluding the effect of $z$, 
the $\alpha_{ro}$--$\alpha_{x \gamma}$ correlation is significant
at the $99.1$ \% level. 
By the same partial correlation analysis, we have investigated
the dependence of the $\alpha_{ro}$--$\alpha_{x \gamma}$
correlation upon the $\gamma$--ray luminosity $L_\gamma$.
The significance, excluding the effect of $L_\gamma$, decreases to the 
97.2 \% level.
Therefore we conclude that the $\alpha_{ro}$--$\alpha_{x \gamma}$ correlation  
is not induced by redshift and it is significant even when considering the 
dependence on $L_\gamma$, which however plays a relatively important role.

These findings coupled with the $\alpha_x$--$\alpha_{\gamma}$ 
relation described in $\S$ 4.1 points to an underlying physical mechanism 
linking the two classes of objects.

\section{Discussion}

\subsection{One or two electron population?}

The overall $\nu F(\nu)$ spectrum of blazars shows two peaks:
the low energy one at IR/soft X--ray energies, and the high energy
one peaking in the MeV/GeV range.
These two peaks can be due to a $single$ population of electrons,
emitting synchrotron and inverse Compton radiation, or can be the
result of the emission of $two$ different electron populations,
emitting both by the synchrotron process, as in the `proton blazar' 
model of Mannheim (1993).
In the former case (one population), the two frequencies at which
the overall emission peaks are obviously related, being produced
by the same electrons.
This holds whatever is the origin of the seed photons to be
upscattered at higher energies.
If this is the case, then one expects correlations between 
fluxes and slopes in different energy bands, of the kind 
found in the present paper, between the slopes at X--rays and $\gamma$--ray
energies, and between the broad band spectral indices connecting
the radio with the optical, and the X--ray with the $\gamma$--rays.

To illustrate our findings, we plot in Fig. 4 the overall
spectra of two blazars: the low redshift BL Lacertae object Mkn 421,
and the high redshift low polarized quasar 0836+710.
As can be seen, both peaks (synchrotron and inverse Compton peaks)
are at smaller energies for 0836+710, and in this object the $\gamma$--ray
luminosity is more dominant.
We have chosen these two sources because they 
are representative of the two classes and have very different BBEDs.

Note that the X--ray emission of Mkn 421 connects smoothly with
the lower energy part of the spectrum, suggesting that in this source
the X--rays are due to the synchrotron mechanism.
This implies the presence of very energetic 
electrons, whose inverse Compton emission can then account for
the observed emission in the TeV band.
Instead, the BBED of 0836+710 suggests that the synchrotron peak
occurs in the IR part of the spectrum, indicating that the
electrons radiating at this peak have smaller energies than in the 
Mkn 421 case.
As a result, the high energy peak should move to the MeV range. 
The extrapolations of the steep $\gamma$--ray 
and the flat X--ray spectra indeed form a peak at a few MeV.

The found anti--correlation between $\alpha_X$ and $\alpha_\gamma$
can be easily interpreted in the `one electron population' scenario.
Steep X--ray spectra 
smoothly connecting with lower frequencies emission imply
synchrotron radiation and therefore
a high frequency for the `Compton' peak, possibly beyond
the EGRET band is expected. This in turn implies flat $\gamma$--ray slopes.

This scenario can also account for the anti--correlation
between $\alpha_{ro}$ and $\alpha_{x\gamma}$.
As in Mkn 421 (Figure 4) a flat $\alpha_{ro}$ 
occurs when the synchrotron component peaks at frequencies above 
the optical band.
This implies a large X--ray flux, due to synchrotron emission,
and a relatively small 100 MeV flux, since the peak of the $\gamma$--ray
emission is above 100 MeV: as a consequence, flat $\alpha_{ro}$
imply steep $\alpha_{x\gamma}$ indices.

The idea of an energy break moving from X--ray to optical--infrared
frequencies and related to the maximum energy of relativistic electrons
has been suggested by Padovani \& Giommi (1995) in order 
to explain the difference between radio and X--ray selected BL Lac objects.
Indeed this idea, modified to take into account the relation
between the frequency of the `synchrotron peak' ($\nu_{S}$) 
and the bolometric synchrotron luminosity,
can explain the different numbers of X--ray and/or radio bright
BL Lac objects selected from flux limited surveys, their
different redshift distributions and evolution (Fossati et al. 1996).
A continuity of properties among the various classes of blazars 
has been also suggested by Sambruna et al. (1996)
from a multiwavelength analysis of sources (mainly BL Lac objects)
detected with the ROSAT PSPC.
The inclusion of the $\gamma$--ray data available by definition in our 
sample, allowing to study the BBED over a much broader energy range,
confirms and extends these findings (see also Padovani \& Giommi 1995;
Sambruna et al. 1996; Maraschi et al 1995).

In the case of {\it two} electron populations an explanation of
all these findings would require a fine tuning between the extension 
in energies and total energy content of the two distributions.

Futhermore, the coordinated variability of
the optical and $\gamma$-ray flux (Maraschi et al. 1994 and Hartman et
al. 1996 for 3C 279; Wagner et al. 1995 for 1406--076) 
can be naturally accounted for by a {\it ``single''}
population model, while in the case of {\it two } electron distribution
scenario the correlated variability would require a even tighter tuning. 

We then conclude that one electron population is a simpler
explanation of the BBED of $\gamma$--ray bright blazar,
with respect to the proton blazar scenario.
If true, this means that the overall synchrotron behaviour
at IR--UV frequencies (peak energy and variability) 
allows to predict the behaviour at high energies: by monitoring
blazars in the optical UV we therefore monitor
the same electrons making the high energy emission.

Note that the $\alpha_{ro}$--$\alpha_{x \gamma}$ relation
suggests that the more $\gamma$--ray dominated sources are on average 
optically faint and are also expected to have a steep spectrum in the 
infrared--optical range. 
If this is the case, relatively bright $\gamma$--ray sources could then be 
discovered among red FSRQ. We point out that the source 
with the steepest $\alpha_{ro}$ (and the flattest $\alpha_x$) in our sample
(i.e. 0202+149) shows an extremely red optical--infrared spectrum 
without any evidence of reddening (Stickel et al. 1996). 
A selection based on optical colors may provide a starting sample
for future more sensitive $\gamma$--ray missions (i.e. GLAST).

\subsection{Implications on the emission models}

A single population of electrons can produce the entire BBED
of blazars by the synchrotron and the inverse Compton process.
However, as mentioned in the introduction,
we still do not know the origin of the seed photons
for the inverse Compton process.

As noted in Dondi and Ghisellini (1995), the $\gamma$--ray
emission in BL Lacertae objects is less dominant than in FSRQ,
even if it extends to higher energies.
Together with the absence of emission lines, this suggests that
the main radiation mechanism in BL Lacertae objects is the 
synchrotron self--Compton (SSC) process.
The (very approximate) equality of the luminosity of the two components
suggests that the magnetic field is close to equipartition with the
synchrotron radiation energy density.
We then applied a one--zone homogeneous SSC model
to the overall emission of Mkn 421, for which simultaneous data
are available both for a `quiescent' state and a flaring state
(Macomb et al. 1995) with the parameters listed in Table 6.
A similar model has been applied in Ghisellini, Maraschi \& Dondi (1996).
We have assumed that in a spherical source of size $R$
electrons are continuously injected, between $\gamma_{min}$
and $\gamma_{max}$, with a power law distribution function,
corresponding to an injected (intrinsic) luminosity $L_{inj}$ and
a compactess $\ell_{inj}\equiv \sigma_T L_{inj}/(Rmc^3)$.
The steady particle distribution, responsible for the emission,
is found self-consistently, through a continuity equation,
by balancing the injection rate with the cooling rate, and taking
into account Klein Nishina effects, pair production and
adiabatic losses (Ghisellini, 1989).
As can be seen, both states of the source can be reproduced
by varying only the injected luminosity (by a factor 2), and
$\gamma_{max}$ (by almost a factor 3).
Incomplete synchrotron and inverse Compton cooling is responsible for 
the curved spectral energy distribution, even if the injected electron 
distribution function is a single power law in the entire energy range.
As a result, the electrons responsible for most of the emitted
radiation (both in the synchrotron and the inverse Compton peaks)
have Lorentz factor close to $\gamma_p=3\times 10^5$ in the high state,
and 3 times less in the low state.

For 0836+710, we have applied both the SSC model and the
`external Compton' model (EC), as developed by Sikora et al. 
(1994). The input parameters are listed in Table 6.
In this EC scenario, we have assumed that the radiation produced external 
to the jet is dominating the electron cooling, and is distributed in energy 
as a diluted blackbody, peaking in the UV band. 

Note that both models account for the main characteristics of 
the overall emission of 0836+710.
On the basis of the pure BBED, therefore, one cannot discriminate
among the two models. 
Other information, such as the broad line luminosity, and
the variability pattern in the IR--UV band with respect to the 
$\gamma$--ray band, is necessary in order to select which are the
dominant seed photons for the inverse Compton process.

In both models for 0836+710, the injected electron distribution function
must have a low energy cut--off, resulting in a double power law
steady particle distribution. 
The curved synchrotron and inverse Compton spectra
are due to the form of the injection function, and not
to the incomplete cooling of the electrons.
In this case most of the emission at both the synchrotron and Compton
peaks is produced by  electrons with Lorentz factors close to 
$\gamma_{min}$.
The equilibrium steady particle distribution is in fact a double
power law, with a break at $\gamma_p=\gamma_{min}$. 
Its value, both for the SSC and the EC case,
is lower than the corresponding $\gamma_p$ in Mkn 421.

The main difference between the SSC and the EC models for 0836+710
concerns the value of the magnetic field,
which is close to equipartition with the synchrotron radiation
energy density in the EC case, and much weaker in the SSC case.

Despite the fact that both models can fit the data of 0836+710,
we can nevertheless draw the conclusion that the different
BBEDs of Mkn 421 and 0836+710 are mainly due to a difference
in the value of their typical electron energies $\gamma_p$.
A distribution of $\gamma_p$, therefore, could well be responsible
for the observed variety of BBEDs of blazars.

\subsection{BL Lacertae objects versus FSRQ}

As extensively discussed in the previous sections 
BL Lacertae objects differ from FSRQ as: 
i) they tend to have steeper X--ray 
and flatter $\gamma$--ray spectra; ii) they separate from FSRQ in 
the $\alpha_{ro}$--$\alpha_{x\gamma}$ plane.
Moreover Dondi \& Ghisellini (1995) found that:
iii) the ratio between the $\gamma$--ray and optical
luminosity, $L\gamma/L_o$, which can be considered as a measure of
the `$\gamma$--ray dominance', is of the order of unity in BL Lacertae objects, 
and much greater in FSRQ.
We confirm this result with our larger sample, deriving 
$<L_\gamma/L_o>\simeq 1$ for BL Lacertae objects and 
$<L_\gamma/L_o>\simeq 30$ for FSRQ.

We have already argued that points i) and ii) can be explained
if the overall spectrum is produced by a single population of electrons
whose distribution is characterized by a break energy $\gamma_p$,
which is, on average, smaller in FSRQ with respect to BL Lacertae objects.
Point iii) above suggests that the value of $\gamma_p$ may be related
to the ratio of the Compton to synchrotron luminosity: the smaller 
$\gamma_p$, the larger the Compton dominance.
This can also explain the found dependence of the  
$\alpha_{ro}$--$\alpha_{x\gamma}$ correlation upon $L_\gamma$.

We speculate that this may be related to the electron Compton cooling rate.
In less compact sources, with no important emission lines, the electron 
cooling time is long.
Large values of $\gamma_p$ are then possible as a result of a competition
between acceleration and cooling rates, or, alternatively, between
cooling and adiabatic expansion (or particle escape) rates.

The fact that BL Lacertae objects have no or weak emission lines
then suggests that one key parameter to explain the BBED of all
objects is the relative importance of the broad emission lines.
If present, they can contribute significantly to the Compton cooling,
resulting in small value of $\gamma_p$ and a dominating $\gamma$--ray emission.
If absent, the $\gamma$--ray emission is instead ruled by the
synchrotron radiation energy density, with a relatively lower 
$\gamma$--ray dominance and larger $\gamma_p$.

We stress that this is only one of the possible ideas trying
to explain the overall energy distribution in blazars, 
but has the advantage, besides its semplicity, to unify 
the spectral properties of all blazars on the basis of one
parameter: the dominant Compton cooling agent.
We will investigate this idea in a future paper.

\section{Conclusions}

We have analysed ROSAT PSPC data for 27 bright EGRET and Whipple
sources,
X--ray fluxes for further 26 sources have been retrieved from
the ROSAT All Sky Survey and from other X--ray missions.
The main conclusions can be summarized as follow:

(1) The measured X--ray spectra show a very broad distribution with energy 
indices in the range $\alpha \sim -0.3-2.3$.
In the soft $\sim$ 0.1--2.0 keV X--ray band the mean spectral slope 
of BL Lac objects is significantly steeper than the mean slope
of FSRQ. The reverse is true in the $\gamma$--ray band where FSRQ 
spectra are steeper than BL Lac ones.

(2) Significant anti--correlations have been found between 
$\alpha_x$ and $\alpha_{\gamma}$ and $\alpha_{ro}$ and $\alpha_{x \gamma}$.
The distribution of BL Lacs and FSRQ in these two diagrams (Figures 2 and 3) 
suggests that the two classes of objects represent the extremes 
of a continuous distribution rather than two distinct populations.
These correlations can be explained if the same relativistic 
electrons are responsible for both the radio to optical and the 
hard X to $\gamma$--ray emission, via, respectively, the synchrotron 
and the inverse Compton processes.

(3) We suggest that the important parameter in describing the observed 
BBEDs is the synchrotron peak energy [in a $\nu-\nu F(\nu)$ plot], 
greater in BL Lac objects than in FSRQ.

(4) We speculate that the proposed distribution in the values
of the peak energy of the synchrotron component may be due
to the different amount of radiative cooling suffered by
the relativistic electrons. 
More severe radiative cooling would result in a small value of the 
synchrotron peak energy and a more dominating $\gamma$--ray emission.

(5) It is predicted that relatively bright $\gamma$--ray sources have
red optical colors (or viceversa). Future missions in the $\gamma$--ray
(i.e. GLAST) and near infrared optical spectrophotometry should be
able to test this hypothesis.

\acknowledgments

We are grateful to H. Liszt who kindly provided 
recent results on CO absorption values, to O. Bendinelli for 
useful discussions on statistical analysis tests, to G. Zamorani,
G.G.C. Palumbo and M. Cappi for a careful reading of the manuscript.
An anonymous referee is thanked for several suggestions resulting in a 
substantial improvement of the original version of the paper.
This work has been partially supported by the Italian Space Agency
(ASI--94--RS--96). GF acknowledges the Italian MURST for financial support.
This research has made use of the NASA/IPAC Extragalactic Database
(NED) which is operating by the Jet Propulsion Laboratory, California
Institute of Technology under contract with the National Areonautic
and Space Administration.

\newpage
\figcaption[] 
{Histograms of the X--ray and $\gamma$--ray spectral
indices for BL Lacs and FSRQ. 
The adopted values for the X--ray 
spectral indices are those listed in Table 5.}

\figcaption[]
{$\gamma$--ray vs X--ray spectral index for the sources in our sample. 
The adopted values for the X--ray spectral indices are those listed in 
Table 5. Filled symbols refer to BL Lac objects, empty symbols refer to FSRQ.}

\figcaption[]
{Correlation between the 
broad band spectral indices  $\alpha_{ro}$ and
$\alpha_{x\gamma}$ for the objects in our sample.
Filled symbols refer to BL Lac objects, empty symbols refer to FSRQ. 
X--ray bright BL Lacs (XBL) have been highlighted with their name,
they cluster in the lower-right part of the diagram. 
We have also labelled the object with the steepest $\alpha_{RO}$.
For 1652+398 and 2344+514, which have not been detected
by EGRET, we have assumed a $\gamma$--ray flux of
0.8$\times 10^{-7}$ photons cm$^{-2}$ s$^{-1}$, approximately
equal to the sensitivity threshold of EGRET. }

\figcaption[]
{Two examples of BBEDs illustrating the difference
among BL Lac objects and FSRQ. The data have been collected from the
literature and the references are reported in Table 5.
The overall spectrum of 0836+710 is fitted with the SSC model
(solid line) and the EC (external Compton) model (dashed line),
as explained in the text.
In both cases a simple one--zone homogeneous sphere has been assumed.
Input parameters for both models are listed in Table 6.
In the EC model, relativistic electrons produce, by synchrotron, the mm
to UV emission, while the high energy spectrum is due to inverse Compton
scattering off seed photons produced externally to the jet.
The ``synchrotron" and ``Compton'' peaks are due to electrons with
Lorentz factors $\gamma_p \sim 150$. 
In the SSC model the `synchrotron' and the `Compton' peaks are instead
due to electrons with Lorentz factor $\gamma_p\sim 4\times 10^3$.
The soft X--ray data are from the present work, while the 0.5--10 keV spectral 
parameters have been retrieved from a recent ASCA analysis 
(Comastri et al. 1996).
For Mkn 421, the overall spectra refer to the May 1994 flare
(circles) and to the preceding quiescent period (squares)
(Macomb et al. 1995). 
The filled square corresponds to the ROSAT observation analyzed in this paper.
The solid and dot-dashed lines refer to one--zone, homogeneous synchrotron
self--Compton models (see also Ghisellini, Maraschi \& Dondi, 1996),
with no contributions from external photons.
Input parameters are listed in Table 6.
The synchrotron and self Compton peaks are due to electrons
of Lorentz factor $\gamma_p \sim 10^5$.
The two different spectra are obtained by varying the number
of emitting electrons (by a factor 3) and their maximum energy
(by a factor 2).}

\end{document}